\documentclass[twocolumn,english,aps,prl,groupedaddress,amsmath,amssymb,footinbib]{revtex4}

\usepackage{mathptmx}
\usepackage[latin9]{inputenc}
\usepackage{prettyref}
\usepackage{graphicx}

\makeatletter

\usepackage{bm}
\usepackage{graphicx}

\makeatother
\begin{document}

\title{Polymerization Force Driven Buckling of Microtubule Bundles Determines
the Wavelength of Patterns Formed in Tubulin Solutions}

\author{Yongxing Guo}

\author{Yifeng Liu}

\author{Jay X. Tang}

\author{James M. Valles, Jr.}

\affiliation{Physics Department, Brown University, Providence, RI 02912}

\date{\today{}}

\begin{abstract}
We present a model for the spontaneous formation of a   striated pattern
in polymerizing microtubule solutions. It describes the buckling of
a single microtubule (MT) bundle within an elastic network formed
by other similarly aligned and buckling bundles and unaligned MTs.
Phase contrast and polarization microscopy studies of the temporal
evolution of the pattern imply that the polymerization of MTs within
the bundles creates the driving compressional force. Using the measured
rate of buckling, the established MT force-velocity curve and the
pattern wavelength, we obtain reasonable estimates for the MT bundle
bending rigidity and the elastic constant of the network. The analysis
implies that the bundles buckle as solid rods.
\end{abstract}
\maketitle
\newpage

\begin{figure}
\begin{centering}
\includegraphics{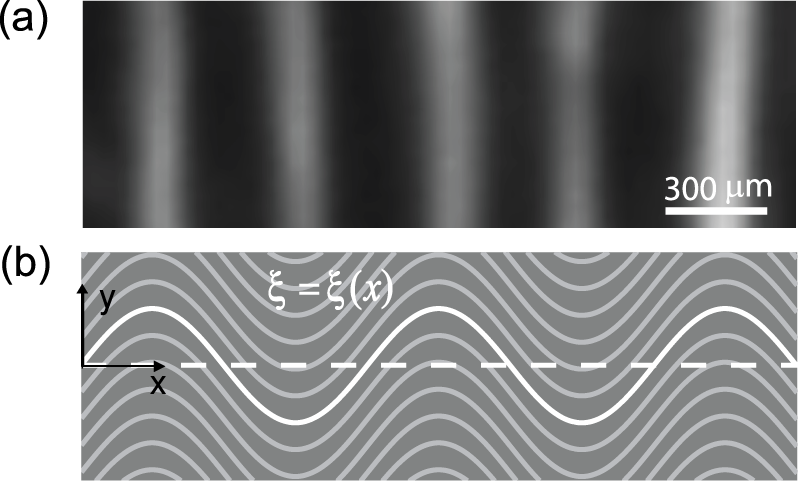}
\par\end{centering}

\caption{\label{fig:Model}Image of a MT birefringent pattern and a sketch
of the mechanical buckling model. (a) Striped birefringent pattern\,\cite{Prl-phase-sample}.
The image was taken between crossed polarizers with the polarization
directions at 45$^{\circ}$ with respect to the $x$ axis. (b) Schematic
drawing of buckled MT bundles surrounded by an elastic MT network
(gray background). The white dashed line depicts the central bundle
before buckling. The white sinusoidal curve depicts the elongated
bundle after buckling and the gray sinusoidal curves represent the
neighboring MT bundles. $\xi(x)$ is the transverse displacement for
the central bundle.}

\end{figure}

Microtubules (MTs), a major component of the eukaryotic cytoskeleton~\cite{Microtubule_Polymerization_Dynamics1997},
can form various structures and patterns. For example, \emph{in}  \emph{vivo,}
MTs organize into the spindles and asters essential for mitosis~\cite{cellmovement}
and the parallel arrays and stripes necessary for  directing early
processes in embryogenesis~\cite{arrays,drosophila}. Many \emph{in
vitro} studies of MT organization have been performed in order to
elucidate the mechanisms underlying the formation of these structures~\cite{Hitt,Tabonyscience,constrained,1998-Walczak-Microtubule}.
Of particular relevance here are the striped birefringent patterns
{[}Fig.~\ref{fig:Model}(a)], which spontaneously form from polymerizing
a purified tubulin solution without motor proteins or MT associated
proteins. Hitt \emph{et al.} attributed these patterns to the formation
of nematic liquid crystalline domains~\cite{Hitt}. Tabony \emph{et
al.}, on the other hand, proposed that a reaction-diffusion based
mechanism drives the formation of MT stripes~\cite{Tabonyscience}.
Our recent investigations imply a starkly different scenario in which
the local MT alignment into wave-like structures occurs through a
 collective process of MT bundling and buckling~\cite{YF_YX_JM_JT_PNAS_2006}.
MTs  that are aligned by a static magnetic field~\cite{Bras-Magnetic,tabony_magnetic_field}
or convective flow~\cite{YF_YX_JM_JT_PNAS_2006} during the initial
stage of polymerization spontaneously form bundles in tubulin solutions
with concentrations of  a few mg/ml. These bundles elongate and buckle
in coordination with neighboring bundles into a wave-like shape. The
nesting of the buckled bundles can quantitatively account for the
 MT density and orientation variations  leading to the striped birefringent
pattern~\cite{YF_YX_JM_JT_PNAS_2006}.  We proposed that  a compressional
force is generated by MT polymerization  occurring uniformly along
the bundle contours.  The buckling  wavelength is controlled by the
bending rigidity of the bundles and the elasticity of the background
network of MTs. This interesting initial assessment calls for  further
investigation  of the microscopic picture of the bundle elongation,
the MT buckling force and the buckling mode selection mechanism.

Here we present a mechanical model for the process  in addition to
new experimental data on the time evolution of the bundle contour
length and solution birefringence that provide direct support  for
 the validity of the model. The model considers the instability of
a single MT bundle under a compressional force, embedded in an elastic
network formed by both bundled and dispersed MTs. Time lapse phase
contrast and quantitative polarized light microscopy imply that MT
polymerization within the bundles provides the compressional force.
Specifically, they reveal that the bundles elongate uniformly along
their contours while maintaining  constant radii consistent with growth
through the elongation of the individual MTs comprising them. We make
predictions for the characteristic buckling wavelength using the bundle
bending rigidity and the critical buckling force estimated from the
measured MT force-velocity curve.  The measured wavelength of about
$600\,\mu\textrm{m}$ implies that the bundles bend as solid rods.

We envision initially the microtubule solution to consist of an array
of straight and parallel bundles aligned along the $x$ axis and embedded
in a  network composed of dispersed MTs as  in Fig.~\ref{fig:Model}(b).
All of the bundles experience a similar compressional force that
grows to a critical value,  causing them to buckle. To describe the
buckling, we consider a single bundle in the center of the sample
and characterize  its interaction with  the  network using a single
elastic constant, $\alpha$,  such that $\alpha\xi(x)$ is the elastic
restoring force exerted by the network on the bundle per unit length.
 Treating the bundle as a  rod with a bending rigidity, $K$, under
a uniform compressional force, $F$,  the force balance in the $y$
direction  at the onset of the buckling is given by~\cite{MT-buckling,dynamic_buckling_Gladden,Theory_of_Elasticity}

\begin{equation}
K\frac{\partial^{4}\xi(x)}{\partial x^{4}}+\frac{\partial}{\partial x}[F\frac{\partial\xi(x)}{\partial x}]+\alpha\xi(x)=0\label{eq:EOM}\end{equation}

Performing a standard normal mode stability analysis of Eq.~\prettyref{eq:EOM}
using $\xi(x)\propto e^{ikx}$ yields  a relation between the angular
wavenumber, $k$, and the compressional force, $F=\alpha/k^{2}+Kk^{2}$,
which suggests a minimum  or critical compressional force $F_{c}$
for a buckling solution.  The critical compressional force is $F_{c}=2\sqrt{K\alpha}$,
and the characteristic wavelength is

\begin{equation}
\lambda_{c}=2\pi/k=\pi\sqrt{8K/F_{c}}=2\pi\sqrt[4]{K/\alpha}\label{eq:characteristic_wavelength}\end{equation}

The resultant characteristic wavelength {[}Eq.~\prettyref{eq:characteristic_wavelength}]
agrees with the prediction for $\lambda_{c}$ based on energy minimization~\cite{YF_YX_JM_JT_PNAS_2006}.
This model predicts buckling in a higher mode than the fundamental
one as in classic Euler buckling.

In agreement with experiments, this model implies that the orientation
of MT bundles in a striped sample varies continuously in space~\cite{YF_YX_JM_JT_PNAS_2006}.
In contrast,  previous models had suggested that discrete and alternate
angular orientations of the MTs formed the striated patterns~\cite{concentration}.
 In addition, the weak dependence of the buckling wavelength on the
mechanical parameters is consistent with the small variations in both
the observed  buckling wavelength   across a single macroscopic sample
and the patterns formed under different conditions (for example,
samples with different tubulin concentrations and samples in containers
with different size.).

\begin{figure}
\begin{centering}
\includegraphics{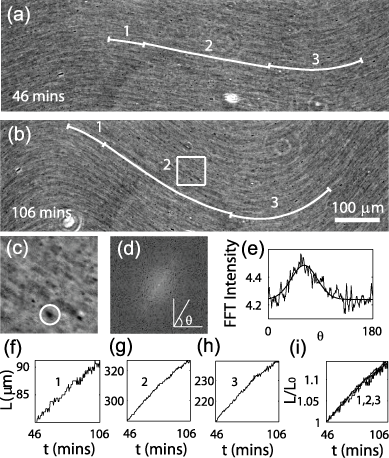}
\par\end{centering}

\caption{\label{fig:Phase}Illustration and measurements of the uniform elongation
of MT bundles\,\cite{Prl-phase-sample}. (a,b) Phase contrast images
of a sample region, show progression of the pattern over one hour.
MT bundles are discerned by the thin striations. The image contrast
is enhanced for better visualization. Segments 1 through 3 are adjacent
pieces of a contour followed by bundles. The segment ends are defined
by fiducial marks. (c) Magnified view of the region denoted by the
white box in (b), showing an encircled fiducial mark. (d) Fast Fourier
Transform (FFT) of (c). (e) The radially averaged FFT intensity plotted
versus the azimuthal angle $\theta$ and fit using a Gaussian function.
The local bundle orientation is orthogonal to the angle at which the
Gaussian fit peaks. (f-h) Length of segments 1 (f), 2 (g) and 3 (h)
as a function of time. (i) Lengths of the three segments as functions
of time, normalized to their lengths at 46 minutes. }

\end{figure}

Time lapse phase contrast microscopy  reveals that the MT bundles
elongate uniformly along their contour during buckling, which is consistent
with polymerization occurring uniformly along the bundles. The elongation
 is illustrated in the  phase images Fig.~\ref{fig:Phase}(a) and
Fig.~\ref{fig:Phase}(b), showing  a fixed region taken 12 and 100
minutes after polymerization initiation, respectively. The three white
curves in each image are computer generated traces of bundle contours
that extend between selected fiducial marks. The fiducial marks are
 visible as dark spots in the images. To generate the white curves,
we presumed that the bundles followed the striations in the images
and traced the stripes  between the fiducial marks, whose positions
were tracked using the MetaMorph imaging software (Universal Imaging,
West Chester, PA). Specifically, we determined the local striation
orientation at each pixel by calculating a Fast Fourier Transform
(FFT) of the area around the pixel, shown, for example,  in Fig.~\ref{fig:Phase}(c).
The FFT appeared as an elongated spot oriented perpendicular to the
striation direction {[}Fig.~\ref{fig:Phase}(d)]. The radially integrated
FFT intensity has a peak at a specific azimuthal angle {[}Fig.~\ref{fig:Phase}(d)]
that is perpendicular to the striation orientation. In this way, the
lengths of three segments along a MT bundle  were recorded every 30
seconds and plotted in Fig.~\ref{fig:Phase}(f), (g) and (h). The
normalized lengths of these three segments grew at nearly the same,
constant rate, shown in Fig.~\ref{fig:Phase}(i), implying that the
MT bundles elongate uniformly along their contour  instead of growing
solely at their ends. It further suggests that the bundles elongate
through polymerization of their constituent MTs, which start and end
at random places along a bundle. The uniform growth of all MTs within
the bundle justifies a uniform elongation rate and the use of a uniform
compressional force throughout the bundle in the mechanical model,
giving rise to the sinusoidal $\xi(x)$ over the entire pattern.

\begin{figure}
\begin{centering}
\includegraphics{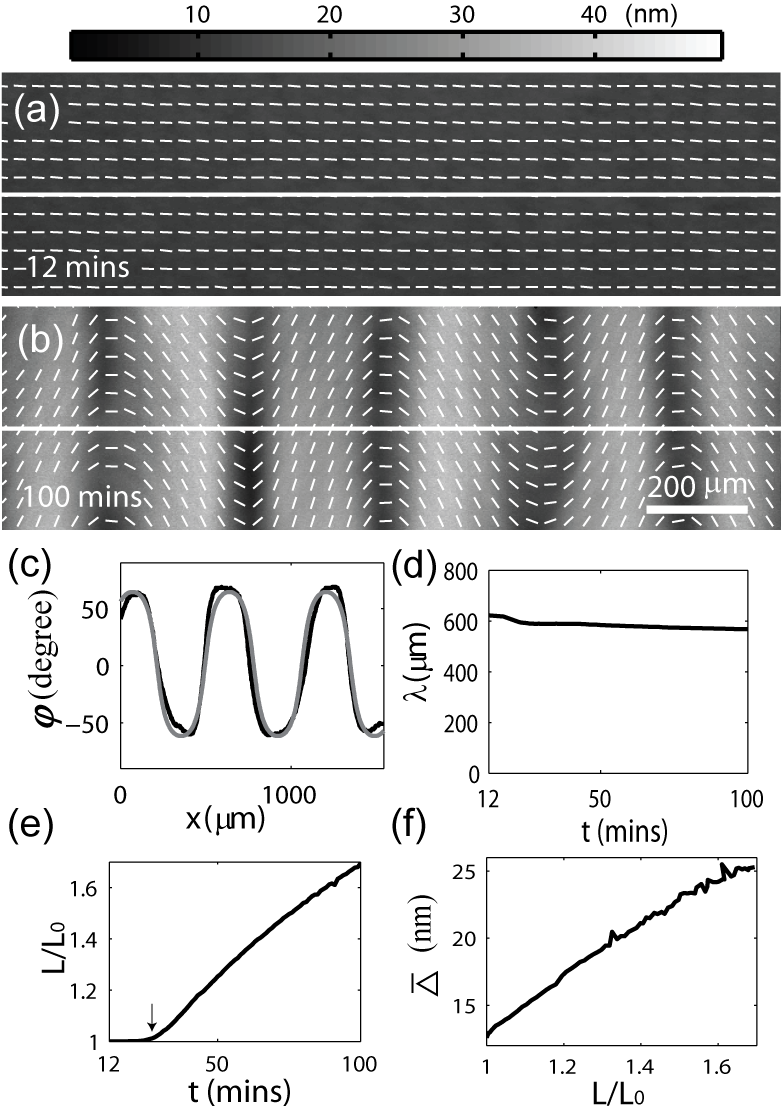}
\par\end{centering}

\caption{\label{fig:Retardance}Time evolution of a MT pattern obtained by
measuring the retardance and slow axis of the sample using a PolScope
imaging system \,\cite{Prl-retardance-sample}. (a,b) Retardance
images of a sample region at 12~and 100~min of self-organization,
respectively. The gray bar shows the retardance magnitude scale and
the white pins provide the slow axis orientation. The straight white
lines represent the slow axis line scan position. (c) Slow axis line
scan (black) and the fitted slow axis orientation $\varphi(x)=\textrm{atan}[A\frac{2\pi}{\lambda}\textrm{cos}(\frac{2\pi}{\lambda}(x+x_{0}))]$
(gray) at 100 min. (d) The dominant buckling wavelength $\lambda$,
obtained from the fitted shapes of the bundle at individual time points.
(e) The length evolution of the fitted bundle contour. $L_{0}=1544\,\mu\textrm{m}$
is the initial unbuckled length of the bundle. The segment before
the arrow designates a latent period prior to the onset of the buckling.
(f) The magnitude of the retardance averaged over the white lines
as shown in (a,b) versus the normalized length $L/L_{0}$. }

\end{figure}

Additional  quantitative information  about the microscopic picture
of the buckling is gained through  time-lapse birefringence measurements.
   PolScope (CRI, Cambridge, MA) images, taken sequentially at a fixed
sample region~\cite{Prl-retardance-sample}, yielded the time evolution
at each pixel of  both the retardance ($\Delta\equiv birefringence\times h$,
where $h$ is the sample thickness) and the slow axis direction ($\varphi(x)$,
orientation of MT bundles)~\cite{oldenbourg_MT_birefringence}.   Two
representative PolScope images of a single region taken at different
stages of self-organization are shown in Fig.~\ref{fig:Retardance}(a)
and \ref{fig:Retardance}(b).  The slow axis variation, $\varphi(x)$,
along the white lines in Fig.~\ref{fig:Retardance}(a) and \ref{fig:Retardance}(b)
can be fit to $\varphi(x)=\textrm{atan}[A\frac{2\pi}{\lambda}\textrm{cos}(\frac{2\pi}{\lambda}(x+x_{0}))]$,
indicating that the bundle follows $\xi(x)=A\sin(\frac{2\pi}{\lambda}(x+x_{0}))$
with a single wavelength $\lambda$, buckling amplitude $A$, and
offset $x_{0}$ {[}Fig.~\ref{fig:Retardance}(c)]. The resultant
wavelength, $\lambda\approx600\,\mu\textrm{m}$, is plotted in Fig.~\ref{fig:Retardance}(d).
 The normalized contour length calculated from the fits, $L(t)/L_{0}$,
 grew nearly linearly with time at a normalized rate of $\dot{L}(t)/L_{0}\approx1$\,\%
per min {[}Fig.~\ref{fig:Retardance}(e)].  Simultaneously, the retardance
magnitude averaged over the white line in Fig.~\ref{fig:Retardance}(a)
increased roughly in proportion to $L(t)/L_{0}$ {[}Fig.~\ref{fig:Retardance}(f)].
Based on the nesting model we proposed earlier and assuming that neighboring
MT bundles do not coalesce, the average retardance goes as $\overline{\Delta}(t)\sim\delta\times n(t)L(t)/L_{0}$~\cite{YF_YX_JM_JT_PNAS_2006,oldenbourg_MT_birefringence},
where $n(t)$ is the number of MTs in the cross section of a bundle
and $\delta$ is the retardance  of a single MT. Therefore, the linear
relation  between  $\overline{\Delta}(t)$ and  $L(t)/L_{0}$ implies
that $n(t)$  remains constant throughout buckling. Thus, the elongation
of MT bundles occurs through the polymerization of MTs within the
bundles and does not involve the incorporation of new MTs to existing
bundles.

\begin{figure}
\begin{centering}
\includegraphics{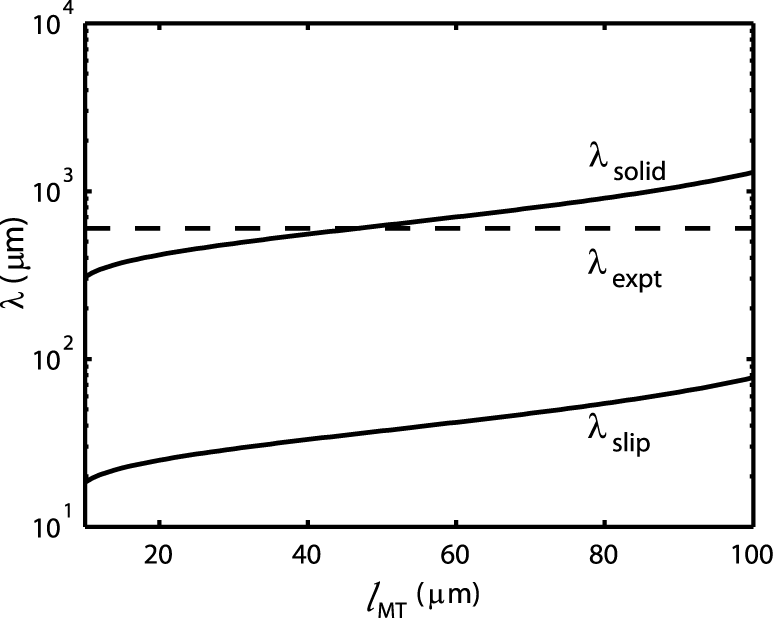}
\par\end{centering}

\caption{\label{fig:ModelSelection} Theoretically calculated wavelength ($\lambda$)
as a function of the average length of MTs ($l_{\textrm{MT}}$) inside
the bundle at the onset of buckling. In the solid model $\lambda_{\textrm{solid}}=\pi\sqrt{8nK_{\textrm{MT}}/f(v(l_{\textrm{MT}}))}$,
and in the slip model $\lambda_{\textrm{slip}}=\pi\sqrt{8K_{\textrm{MT}}/f(v(l_{\textrm{MT}}))}$).
$\lambda_{\textrm{expt}}$ is the experimentally observed buckling
wavelength (dashed line).}

\end{figure}

With the above observations and model, we can quantitatively characterize
the elastic properties of the bundle ($K$) and network ($\alpha$).
We begin with the implications of the measured wavelength $\lambda$.
In order to predict $\lambda$ from the mechanical buckling model,
we need to estimate $K$ and $F$ {[}Eq.~\prettyref{eq:characteristic_wavelength}].
Two limits exist for $K$. If tight packing (solid model) of the MTs
 inside the bundle is assumed, then $K_{\textrm{solid}}=n^{2}K_{\textrm{MT}}$,
where $K_{\textrm{MT}}\approx3.4\times10^{-23}\, N\cdot m^{2}$ is
the  bending rigidity of a single MT~\cite{buckling,2005-Tuszynski-MT-elasticity}.
If MTs  slide freely inside the bundle, then $K_{\textrm{slip}}=nK_{\textrm{MT}}$.
We employ the measured force-velocity relation, $f(v)=C_{1}\ln[C_{2}/(v+C_{3})]$
($C_{1}=1.89\, pN$, $C_{2}=1.13\,\mu\textrm{m}/\textrm{min}$ and
$C_{3}=-0.08\,\mu\textrm{m}/\textrm{min}$~\cite{buckling}), for
a single MT and presume $F=nf(v)$, where $v$ is the average elongation
rate of individual MT inside the bundle. Writing the average length
of MTs inside the bundle  as $l_{\textrm{MT}}$, the elongation rate
of a single MT is then approximately $v(l_{\textrm{MT}})=l_{\textrm{MT}}\times\dot{L}(t)/L_{0}$.
Using the models for $K$, $F$ and Eq.~\prettyref{eq:characteristic_wavelength},
we derive predictions of $\lambda$ for both the solid model, $\lambda_{\textrm{solid}}=\pi\sqrt{8nK_{\textrm{MT}}/f(v(l_{\textrm{MT}}))}$,
and the slip model, $\lambda_{\textrm{slip}}=\pi\sqrt{8K_{\textrm{MT}}/f(v(l_{\textrm{MT}}))}$.
Each depends on $l_{\textrm{MT}}$ and $n$. Using $n=280$~\cite{YF_YX_JM_JT_PNAS_2006},
we plot the wavelength over a reasonable range of individual MT lengths
(\cite{Microtubule_Polymerization_Dynamics1997}) in Fig.~\ref{fig:ModelSelection}.
The solid model for $K$ appears much more reasonable than the slip
model. The fact that $K$ depends quadratically on $n$ in our system
suggests that MTs are fully coupled (acting like a solid material)
inside the bundle, similar to the behavior of F-actin bundles held
together through depletion forces~\cite{2006-Claessens-Bundle-Stiffness}.
The bundling of initially aligned MTs can be attributed to the depletion
force induced by unpolymerized tubulin dimers, oligomers and even
short MTs~\cite{YF_YX_JM_JT_PNAS_2006}.

The conclusion that the bundles bend as solid rods apparently conflicts
with the picture of elongation,  that involves the growth and relative
sliding of individual MTs within the bundles. We speculate that the
explanation involves two distinct time scales: the time for a MT to
come to mechanical equilibrium with its  neighbors following the insertion
of a tubulin dimer to its end, $\tau_{\textrm{mech}}$, and the average
interval between  insertions, $\tau_{\textrm{dimer}}$. In the limit
$\tau_{\textrm{mech}}<\tau_{\textrm{dimer}}$, strong coupling between
the MTs in the bundle can occur leading to the solid rod result. The
opposite limit intuitively leads to weak coupling between  the MTs
within a bundle. We estimate $\tau_{\textrm{dimer}}\approx0.1\,\textrm{s}$
from our data, which seems quite  long compared to the times characterizing
the relative motion of neighboring MTs on the molecular length scales
 relevant to $\tau_{\textrm{mech}}$. The exact molecular picture,
 which goes beyond the scope of our model, needs further study.

Using the solid model for $K$, we can calculate the remaining model
parameter, $\alpha$, from Eq.~\prettyref{eq:characteristic_wavelength}:
$\alpha=K_{\textrm{slip}}(2\pi/\lambda_{\textrm{expt}})^{4}\approx0.032\,\textrm{Pa}$.
This value is remarkably small compared to that estimated for a single
MT buckling inside a cell ($\alpha^{*}\approx2700\,\textrm{Pa}$~\cite{MT-buckling}).
We identify two contributors to the difference between $\alpha$ and
$\alpha^{*}$. In general, $\alpha\sim G$, where $G$ is the elastic
shear modulus of the surrounding network.  $G\sim1\,\textrm{Pa}$
in our system~\cite{elastic_shear_modulus}, while $G^{*}\sim1000\,\textrm{Pa}$
for the surrounding cytoskeleton network inside the cell~\cite{2000-Mahaffy-G-gel}.
 The other contributor is the coordination of the buckling of the
MT bundles, which reduces the distortion of the surrounding network,
and thus weakens the effective restoring force and $\alpha$ (analysis
in preparation).

In summary, using   microscopic studies of the temporal evolution
of the striated MT patterns, we show that the polymerization of MTs
within the bundles causes uniform elongation. This in turn creates
the driving compressional force which ultimately causes the MT bundles
to buckle. It is this coordinated buckling  that produces the striped
birefringent pattern.  The proposed mechanical buckling model adequately
describes the buckling process. It  predicts  a critical buckling
force and a characteristic wavelength, which depend on the elasticity
 of the surrounding network and the bending rigidity of the MT bundles.
Combing   the bending rigidity of  MT bundles and the established
MT force-velocity curve with the mechanical model, we obtain a reasonable
estimate for the elastic constant of the network and find that MTs
inside the bundle are fully coupled.

We thank Allan Bower for help in understanding the elastic constant
$\alpha$ and thank   L. Mahadevan and Thomas R. Powers for valuable
discussions. This work  was supported by NASA (NNA04CC57G, NAG3-2882)
and NSF (DMR 0405156, DMR 0605797).


\begin{thebibliography}{24}
\bibitem[{Prl({\natexlab{a}})}]{Prl-retardance-sample}
\bibinfo{note}{The sample was polymerized from a 5 mg/ml tubulin solution (2 mM
  GTP, 3.5\% in molar ratio of Oregon Green conjugated taxol to tubulin dimers,
  100 mM pipes, 1 mM EGTA, 2 mM MgSO$_4$, PH 6.9) in a
  $40\times8\times0.4\,\textrm{mm}^{3}$ glass cuvette which was exposed to 9\,T
  vertical static magnetic field for 5 minutes at 37$^{\circ}$C (the magnetic
  field direction is along the long axis of the cuvette). The cuvette was then
  laid flat on the microscope stage and a coherently buckled area was chosen
  for observation and measurement at 30$^{\circ}$C.}
\bibitem[{Prl({\natexlab{b}})}]{Prl-phase-sample}
\bibinfo{note}{The sample was polymerized from 8 mg/ml tubulin solution (same
  buffer condition as in \,\cite{Prl-retardance-sample}) in a
  $40\times10\times1\,\textrm{mm}^{3}$ quartz cuvette and was subjected to
  convective flow (induced by asymmetrical thermal contacts, with the left and
  bottom surfaces in contact with a 37$^{\circ}$C waterbath-warmed aluminum
  holder and other sides exposed to 30$^{\circ}$C ambient) for the first 9
  minutes. The cuvette was then laid flat on the microscope stage for
  observation and measurement at 30$^{\circ}$C.}
\end{thebibliography}
\end{document}